\documentstyle[aps,epsf]{revtex}
\begin{document}

\draft

\title{
{\bf{Influence of projectile $\alpha$- breakup threshold on complete fusion 
}}
}

\author{
A. Mukherjee$^{*}$, Subinit Roy, M.K. Pradhan, M. Saha Sarkar, P. Basu, 
B. Dasmahapatra 
}

\address{
Saha Institute of Nuclear Physics, 1/AF, Bidhan Nagar, 
Kolkata-700 064, India
}
\author{
T. Bhattacharya, S. Bhattacharya, S.K. Basu 
}

\address{
Variable Energy Cyclotron Centre, 1/AF, Bidhan Nagar, 
Kolkata-700 064, India
}

\author{
A. Chatterjee, V. Tripathi, S. Kailas 
}

\address{
Nuclear Physics Division, Bhabha Atomic Research Centre, 
Mumbai 400 085, India
}

\date{\today}
\maketitle

\begin{abstract}
	Complete fusion excitation functions for $^{11,10}$B+$^{159}$Tb 
have been measured at energies around the respective Coulomb barriers, 
and the existing complete fusion measurements for $^{7}$Li+$^{159}$Tb 
have been extended to higher energies. The measurements show significant 
reduction of complete fusion cross sections at above-barrier energies for 
both the reactions, $^{10}$B+$^{159}$Tb and $^{7}$Li+$^{159}$Tb, when 
compared to those for $^{11}$B+$^{159}$Tb. The comparison shows that the 
extent of suppression of complete fusion cross sections is correlated with 
the $\alpha$-separation energies of the projectiles. Also, the two reactions, 
$^{10}$B+$^{159}$Tb and $^{7}$Li+$^{159}$Tb were found to produce incomplete 
fusion products at energies near the respective Coulomb barriers, with 
the $\alpha$-particle emitting channel being the favoured incomplete 
fusion process in both the cases.
\end{abstract}


\pacs {Keywords: Heavy-ion fusion; Breakup; Fusion suppression} 
\pacs{$^{*}$ Corresponding author. E-mail address: anjali.mukherjee@saha.ac.in}
\vspace{1.0 cm}
	 Recently a resurgence of interest has occurred in investigating 
the effect of breakup of weakly bound projectiles on the fusion mechanism 
\cite{Yo96,Reh98,Ko98,Sig04,Pie04,Dass93,Ha00,Dia02} at energies around the 
barrier. This has primarily been motivated by the present availability of 
radioactive ion beams, some of which exhibit unusual features like
halo/skin structure and large breakup probabilities. A critical 
understanding of the fusion mechanism with radioactive ion beams is very 
significant for the understanding of reactions of astrophysical interest 
and for the production of new nuclei near the drip lines. 
   
         It can be expected that in the fusion studies involving halo nuclei, 
the larger spatial extent of such nuclei may lead to a lowering of the
average fusion barrier, and thus enhance the fusion cross sections over those 
for well-bound nuclei. On the contrary, the halo nuclei can easily breakup 
in the field of the other nucleus, due to their low binding energies, and can 
therefore lead to a loss of flux from the entrance channel thereby reducing 
the fusion cross sections. However, coupled channels calculations \cite{Ha00} 
carried out for the system $^{11}$Be+$^{208}$Pb show that a combination of 
all these effects essentially leads to enhancement of fusion cross sections at 
sub-barrier energies and reduction of fusion cross sections at above-barrier 
energies. Although presently it is possible to investigate reaction mechanisms 
with unstable beams, experimentally such studies are limited 
\cite{Yo96,Reh98,Ko98,Sig04,Pie04} owing to the low intensities of the 
radioactive beams currently available. On the other hand, fusion reactions 
with high intensity weakly bound stable beams which have a significant breakup 
probability may serve to be an important step towards the understanding of the 
influence of breakup on the fusion mechanism. Indeed in the last few years, 
special attention has been paid towards fusion studies at near-barrier energies
using the weakly bound stable projectiles, $^{9}$Be, $^{6}$Li and $^{7}$Li. 

	In fusion with weakly bound projectiles, following the breakup of the 
projectile in the field of the target, one of the fragments may be captured by 
the target, with the other escaping with the beam velocity \cite{Utso83}. 
This process of capture of partial projectile by the target is known as
incomplete fusion (ICF) and has been observed even in reactions with strongly
bound projectiles like $^{12}$C and $^{16}$O, but mostly at higher bombarding 
energies \cite{Zo78,Bar80}. However, the recent observation of ICF cross 
sections at near-barrier energies in fusion with weakly bound projectiles 
\cite{Das99,Sig99,Das02,Das04,Tri02} has made this field even more interesting, 
especially in view of the present availability of the radioactive ion beams.

          In fusion of $^{9}$Be and $^{6,7}$Li with heavy targets like, 
$^{208}$Pb  and $^{209}$Bi \cite{Das99,Sig99,Das02,Das04} substantial 
suppression of complete fusion cross sections  has been observed at energies 
above the respective Coulomb barriers. The complete fusion (CF) products 
correspond to the events where the whole of projectile fuses with the target.  
For medium and light mass systems, owing to the experimental difficulties, 
total (complete+incomplete) fusion cross sections
were measured for the systems like, $^{9}$Be+$^{64}$Zn \cite{Mor00}, 
$^{6,7}$Li+$^{59}$Co \cite{Dia03}, $^{6,7}$Li+$^{12,13}$C \cite{Mu1,Mu2} and 
$^{6,7}$Li+$^{16}$O \cite{Mu3,Mu4}. These measurements show no suppression of 
total fusion cross sections at above-barrier energies. 

        To investigate the effect of breakup on fusion, all the reactions 
studied so far with weakly bound stable beams have been performed using 
$^{9}$Be, $^{6}$Li and $^{7}$Li projectiles that have breakup thresholds 
ranging from 1.45 MeV to 2.45 MeV. Among the stable nuclei, apart from  
the nuclei $^{6,7}$Li and $^{9}$Be, the nucleus $^{10}$B also has a fairly 
low $\alpha$ separation energy of 4.5 MeV. Therefore like $^{6,7}$Li and 
$^{9}$Be, the nucleus $^{10}$B also may be expected to breakup at low 
excitation energies, thereby affecting the fusion mechanism at considerably 
low bombarding energies. Particle-$\gamma$ coincidence measurements carried 
out with 75 MeV $^{10}$B beam and $^{159}$Tb target show substantial 
production of Er nuclei, resulting from the ICF process \cite{Zo78}. In 
this Letter, we present the CF excitation functions for the  
$^{11,10}$B+$^{159}$Tb and $^{7}$Li+$^{159}$Tb reactions, at energies around 
the respective Coulomb barriers. The CF measurements for $^{7}$Li+$^{159}$Tb 
have been extended to energies higher than that reported in the literature 
\cite{Bro75}. The $^{11}$B projectile, with $\alpha$ separation energy of 
8.66 MeV, is expected to behave as a normal strongly bound nucleus. Thus, 
$^{11}$B+$^{159}$Tb was chosen to be the reference strongly bound system. 
A comparison of the CF cross sections for the three systems at 
above-barrier energies allows to study the correlation between the extent of 
CF suppression and the $\alpha$-breakup thresholds of the projectiles. 

           Beams of $^{11,10}$B in the energy range 38$-$72 MeV, and $^{7}$Li 
with energies from 28$-$43 MeV, provided by the 14UD BARC-TIFR Pelletron 
Accelerator Facility at Mumbai, bombarded a self-supporting $^{159}$Tb target 
of thickness 1.50$\pm$0.07 mg/cm$^{2}$. The $\gamma$-rays emitted by the 
evaporation residues (ERs) were detected in an absolute efficiency calibrated 
Compton suppressed clover detector placed at 55$^{o}$ with respect to the 
beam direction. To cross check the measured cross sections, a 125 c.c. HPGe 
detector was also placed at 125$^{o}$ with respect to the beam direction. 
Both online and offline spectra were taken for each exposure. The total charge
of each exposure was measured in a 1m long Faraday cup placed after the
target. The target thickness was determined by measuring the Rutherford
scattering cross sections and also by using the 137.5 keV Coulomb excitation
line of $^{159}$Tb. The thickness of the target obtained from the two methods 
of measurement had excellent agreement. 

           The compound nuclei $^{170}$Yb, $^{169}$Yb and $^{166}$Er, formed
following the fusion reactions $^{11}$B+$^{159}$Tb, $^{10}$B+$^{159}$Tb and 
$^{7}$Li+$^{159}$Tb respectively, are expected to decay predominantly by
neutron evaporation producing ERs which are all well deformed nuclei. This is 
also predicted by the statistical model calculations done using the code 
PACE2 \cite{Gav80}. The CF cross sections in the B induced reactions were 
obtained from the sum of the $3n - 6n$ ER cross sections and for the $^{7}$Li 
induced reaction the same was obtained by summing the $3n - 5n$ ER cross 
sections. 

           For the even-even ERs, except the $3n$ channel ($^{166}$Yb) in the 
reaction $^{10}$B+$^{159}$Tb, the $\gamma$-ray cross sections, $\sigma$(J), 
for various transitions in the ground state rotational band of the relevant 
nucleus were obtained using the measured $\gamma$-ray intensities after 
correcting for the internal conversion. The cross sections for a given 
even-even ER were then extracted from the extrapolated value of the 
$\gamma$-ray cross section at J=0. For the odd-mass nuclei, the cross sections 
were obtained by following the respective radioactive decay. The low lying 
characteristic $\gamma$-rays in the ground state band of the even-even ER 
$^{166}$Yb, corresponding to the $3n$ channel in the 
$^{10}$B+$^{159}$Tb reaction, are almost identical to those in the nucleus 
$^{162}$Er, a probable ICF product in this reaction. So the measured 
cross sections for the $\gamma$-rays corresponding to the $^{166}$Yb nucleus 
will also include the contributions from the $^{162}$Er nucleus. In order to 
estimate the contribution from the $^{166}$Yb ER, and hence extract the 
contribution from the ICF product $^{162}$Er, if any, we adopted the following 
procedure. For the reaction $^{10}$B+$^{159}$Tb, it was assumed that there is 
no significant contribution from the ICF process at energies below the Coulomb 
barrier. So at energies below the Coulomb barrier the measured cross sections 
correspond almost wholly to the $^{166}$Yb ER. At below-barrier energies, the 
ratio, $F$=
$\frac{\sigma(3n/4n)_{^{10}B+^{159}Tb}}{\sigma(3n/4n)_{^{11}B+^{159}Tb}}$ was 
obtained from the measured cross sections, at the same excitation energies for
both the reactions. The cross sections for the $^{166}$Yb ER at above-barrier 
energies were then obtained using the measured $4n$ channel cross sections for 
$^{10}$B+$^{159}$Tb, the factor $F$, and the measured ratios of the cross 
sections for $3n(^{167}$Yb)/$4n(^{166}$Yb) in the $^{11}$B+$^{159}$Tb system at
the same excitations energies as in the $^{10}$B case. By this normalisation it
was assumed that the ratio, $F$=
$\frac{\sigma(3n/4n)_{^{10}B+^{159}Tb}}{\sigma(3n/4n)_{^{11}B+^{159}Tb}}$ is
constant over the whole energy region of measurement, and this constancy was
checked using the statistical model calculations performed using the code
PACE2. The contribution of the ICF product $^{162}$Er was then obtained using
the contribution of $^{166}$Yb, determined as above, and the measured total 
cross sections for the $\gamma$-rays corresponding to $^{166}$Yb (or 
$^{162}$Er). It needs to be mentioned here that the cross sections obtained 
using the data from the clover detector agreed very well with those obtained 
with the HPGe detector. 

      The  CF cross sections for the three reactions $^{11,10}$B+$^{159}$Tb and
$^{7}$Li+$^{159}$Tb were then determined by summing the respective $xn$ channel 
cross sections at each energy. The results are shown in Figs. 1-3. The CF data 
of Broda {\em{et al.}}\cite{Bro75} for the $^{7}$Li+$^{159}$Tb reaction are 
shown by the hollow points in Fig. 3. As expected, the $^{11}$B+$^{159}$Tb 
system behaves as a normal strongly bound system, where no $\gamma-$lines 
following ICF were observed in the spectra over the energy range of present 
measurement. Cross sections for the $3n$ and $4n$ channels in the decay of 
$^{165}$Er, formed following the capture of $^{6}$Li by $^{159}$Tb in the 
$^{10}$B induced reaction, are also shown in Fig. 2. Cross sections for the 
dominant $\alpha2n$ channel, following the $t$ capture by $^{159}$Tb in the 
$^{7}$Li+$^{159}$Tb reaction, are shown in Fig. 3. It needs to be mentioned 
that the ICF cross sections plotted in Figs. 2-3 include contributions from 
breakup fusion and transfer from projectile to the unbound states of the 
target. The errors in the cross sections plotted in Figs. 1-3 are the total 
errors, which include statistical errors and uncertainties in the target 
thickness, efficiency of the detector and the integrated beam current. 

        In the $^{7}$Li+$^{159}$Tb reaction, the contribution from the capture 
of the lighter projectile fragment, $t$, by $^{159}$Tb was found to be the 
dominant ICF contribution, with the contribution from $\alpha$+$^{159}$Tb being
negligibly small. A similar observation was reported for the reaction 
$^{7}$Li+$^{165}$Ho \cite{Tri02} and was explained to be due to the higher 
Coulomb barrier for the $\alpha$- capture compared to the $t$- capture. 
By contrast, in the $^{10}$B+$^{159}$Tb reaction, the $\gamma$-spectra 
showed no lines corresponding to the $\alpha$ (lighter fragment) capture 
by $^{159}$Tb. In this reaction, the only ICF contributions which could be 
observed were due to the capture of $^{6}$Li (heavier fragment) by $^{159}$Tb, 
even though the Coulomb barrier for the $\alpha$- capture is lower than the 
$^{6}$Li- capture by the target. This observation is indeed consistent with 
the corresponding Q-values of the reactions. The Q-value for the 
$^{159}$Tb($^{7}$Li,$\alpha$)$^{162}$Dy reaction is 11.1 MeV, while it is 
$-$3.2 MeV for the ($^{7}$Li,$t$) reaction, indicating that the $\alpha$ 
particle emission is more favoured. In the $^{10}$B induced reaction, the 
Q-value for the $^{159}$Tb($^{10}$B,$\alpha$)$^{165}$Er reaction is 4.6 MeV, 
while it is $-$5.2 MeV for the ($^{10}$B,$^{6}$Li) reaction, indicating 
that $\alpha$ particle emission is more favoured. In fact, in both the cases, 
the favorable ICF channel is where the $\alpha$ particle emission occurs.

	To compare the CF cross sections for the three reactions at 
above-barrier energies, they have been plotted in a reduced scale in Fig. 4. 
In this figure only statistical uncertainties have been plotted, as the 
present measurements for the three systems were taken in one run using the 
same setup. The data of Ref. \cite{Bro75} have been plotted with the errors 
quoted in the article. The figure clearly shows in a model independent way 
that the CF cross sections for $^{10}$B+$^{159}$Tb and $^{7}$Li+$^{159}$Tb are 
suppressed at above-barrier energies compared to those for $^{11}$B+$^{159}$Tb,
with the cross sections for $^{10}$B+$^{159}$Tb being intermediate between 
those for $^{11}$B+$^{159}$Tb and $^{7}$Li+$^{159}$Tb. This observation is 
quite consistent with the $\alpha$-breakup thresholds of the projectiles. 
As discussed earlier, of the three projectiles, $^{11}$B is the most strongly 
bound nucleus with Q$_\alpha$=$-$8.46 MeV and $^{7}$Li is the most weakly 
bound nucleus with Q$_\alpha$=$-$2.47 MeV. The $^{10}$B nucleus has 
Q$_\alpha$=$-$4.5 MeV, intermediate between that of $^{11}$B and $^{7}$Li. 
Thus, lower the $\alpha$- breakup threshold of the projectile, larger is the 
suppression of CF. Moreover, Fig. 4 also shows that the onset of suppression 
depends on the $\alpha$-separation energy of the projectile. Higher the 
breakup threshold, higher is the energy where the suppression starts. This 
perhaps explains why ICF products are observed in strongly bound systems at 
much higher bombarding energies \cite{Zo78,Bar80}. 

       To study the extent of above-barrier fusion suppression in a theoretical
framework, the realistic coupled channels (CC) code CCFULL \cite{Hag99} was 
employed to calculate the total fusion cross sections. It needs to be pointed 
out here that these calculations do not consider couplings to unbound or 
continuum states. Thus the breakup of the projectiles $^{10}$B and $^{7}$Li 
is not included. In CCFULL the number of CC equations is reduced by means of 
the isocentrifugal approximation, and an ingoing-wave boundary condition is 
placed inside the barrier. 

The Aky$\ddot{\mbox{u}}$z-Winther (AW) \cite{Brog81} (bare) potential 
parameters ($V_{0}$, $r_{0}$, and $a$) for the three systems are given in 
Table I. The corresponding uncoupled fusion barrier
parameters ($V_{b}$, $R_{b}$, and ${\hbar}\omega$) are also mentioned in the
table. The CCFULL calculations with the shallow AW potentials lead to 
oscillations of transmission coefficients of high partial waves, especially at 
high energies. To minimize such oscillations, the potential wells for the three
systems were chosen to be deeper so that the ingoing-wave boundary condition is 
correctly applied. The diffuseness parameter was chosen to be $a=0.85$ fm for
all the three systems, following the systematic trend of high diffuseness 
required to fit the high energy part of the fusion excitation functions 
\cite{New1}. The radius parameter had to be changed accordingly. For 
$^{11}$B+$^{159}$Tb, with $a=0.85$ fm, $V_{0}$ and $r_{0}$ were varied so as
to fit the high-energy cross sections ($>$200 mb) \cite{Mu5}. This modified
potential for CC calculations are given in Table I. For 
$^{10}$B+$^{159}$Tb, the same potential parameters were used in the CC 
calculations as they are very nearby systems. The similarity in the potential 
parameters for the two systems can also be observed in the AW potentials. In 
the case of $^{7}$Li+$^{159}$Tb, keeping $a$ fixed at 0.85 fm, $V_{0}$ and 
$r_{0}$ were varied so that the corresponding one-dimensional barrier 
penetration model (1-D BPM) cross sections for the $^{7}$Li+$^{159}$Tb agree 
with those with the AW potential parameters at higher energies. The 1-D BPM 
calculations were done using the code CCFULL, in the no coupling limit.

          In CCFULL, the effects of deformation are calculated by coupling to 
the ground state rotational band of the deformed target nucleus. The target 
$^{159}$Tb is a well-deformed nucleus with an unpaired valence proton.  
This last valence proton particle (or proton hole) can be 
expected to couple with the $0^{+}$, $2^{+}$, $4^{+}$, .... rotational states 
present in the neighbouring even-even rotational nucleus $^{158}$Gd (or 
$^{160}$Dy) to build up the low-lying rotational states of $^{159}$Tb. 
To remain within the model space of CCFULL, the excitation energies and
deformation parameters for $^{159}$Tb were taken to be the averages of the
corresponding values for the neighbouring even-even nuclei $^{158}$Gd and
$^{160}$Dy \cite{Mu6}. The ground state rotational band of the corresponding
average spectrum, ($\beta_{2}$=0.344 \cite{Ram87} and $\beta_{4}$=+0.062
\cite{Mol95}) upto $12^{+}$, were included in the calculations \cite{Mu5}.
Projectile excitations were not considered in any of the cases. The results 
of the 1-D BPM are shown by the dot-dot-dashed lines in Figs. 1-3. The CC 
calculations for $^{11}$B+$^{159}$Tb are shown by the solid lines in Fig. 1 
and those for $^{10}$B+$^{159}$Tb and $^{7}$Li+$^{159}$Tb are shown by the 
dashed lines in Figs. 2-3.  

The CC calculations are found to be in reasonably good agreement with the  
measured fusion cross sections for the $^{11}$B+$^{159}$Tb. But for 
$^{10}$B+$^{159}$Tb and $^{7}$Li+$^{159}$Tb, at above-barrier energies the 
measured CF cross sections lie below the calculated cross sections. 
Using the data above E$_{c.m.}\sim$ 45 MeV for $^{10}$B+$^{159}$Tb and
those above E$_{c.m.}\sim$ 25 MeV for $^{7}$Li+$^{159}$Tb, the measured CF 
cross sections for the two systems are found to be, respectively, $\sim 86\%$ 
and $\sim 74\%$ of the theoretical predictions. This observation is consistent 
with the 
aforementioned model-independent comparison of the three systems. The CC 
calculations for $^{10}$B+$^{159}$Tb and $^{7}$Li+$^{159}$Tb when scaled by the
factors 0.86 and 0.74 respectively, are shown by the solid curves in Figs. 2-3. 
A suppression factor of $\sim$0.74 was also obtained for other $^{7}$Li induced 
reactions, like $^{7}$Li+$^{209}$Bi \cite{Das02,Das04} and $^{7}$Li+$^{165}$Ho
\cite{Tri02}.

      In summary, the CF excitation functions for the three reactions, 
$^{11,10}$B+$^{159}$Tb and $^{7}$Li+$^{159}$Tb have been measured. The CF
cross sections for $^{10}$B+$^{159}$Tb and $^{7}$Li+$^{159}$Tb show 
suppressions of $\sim 86\%$ and $\sim 74\%$ respectively. The extent of this 
suppression is found to be correlated with the $\alpha$-separation energies of 
the projectiles. Besides, it has also been observed that fusion with a 
projectile having a higher $\alpha$-breakup threshold results in the onset of 
CF suppression at a higher bombarding energy. Both $^{10}$B+$^{159}$Tb and 
$^{7}$Li+$^{159}$Tb reactions were found to produce ICF products at energies 
near the respective Coulomb barriers, with the $\alpha$-particle emitting 
channel being the favoured ICF process in both the cases. The present study 
with weakly bound stable nuclei and more such in the near future will perhaps, 
lead to a deeper systematic understanding of the effect of very weak binding 
of the unstable nuclei on the fusion process.
\\
\\        
\\
\\
{\bf{Acknowledgements}}

We would like to thank Prof. H.C. Jain for providing the clover detector.
We also thank Mr. P. Das for preparing the target and the accelerator staff at 
the BARC-TIFR Pelletron Facility, Mumbai, for their untiring efforts in 
delivering the beams. 


\begin{table}
\begin{center}
\caption{The parameters of the AW potential and of the modified potential
used for the CC calculations (see text). Also shown are the corresponding 
derived uncoupled barrier heights $V_{b}$, radii $R_{b}$ and curvatures 
$\hbar\omega$.}
\begin{tabular}{lccccccc}
                                                                                
System &Potential& $V_{0}$ & $r_{0}$ & $a$ & $V_{b}$ & $R_{b}$ & $\hbar\omega$  \\
       &         & (MeV) & (fm) & (fm) & (MeV) & (fm) & (MeV)                   \\
\hline
$^{11}$B+$^{159}$Tb &AW& 54.54 & 1.18 & 0.64 & 40.34 & 10.89 & 4.42 \\
                    &CC& 140 & 1.01 & 0.85 & 39.72 & 10.84 & 3.87 \\
                    &  &     &      &      &       &       &      \\
$^{10}$B+$^{159}$Tb &AW& 54.54 & 1.11 & 0.64 & 40.71 & 10.79 & 4.68 \\
                    &CC& 140 & 1.01 & 0.85 & 40.00 & 10.75 & 4.08 \\
                    &  &     &      &      &       &       &      \\
$^{7}$Li+$^{159}$Tb &AW& 46.43 & 1.18 & 0.62 & 24.70 & 10.69 & 4.48 \\
                    &CC& 132 & 0.98 & 0.85 & 24.17 & 10.68 & 3.81 \\
\end{tabular}
\label{Table I}
\end{center}
\end{table}

{\bf{FIGURE CAPTIONS}}\\
\\
\\
Fig.1.  Fusion excitation function for the $^{11}$B+$^{159}$Tb system. 
The dot-dot-dashed and the solid lines are the uncoupled and coupled channels 
calculations performed with the code CCFULL.\\
\\
Fig.2.  Complete and incomplete fusion cross sections (circle and 
triangle) as a function of the centre-of-mass energy for the 
$^{10}$B+$^{159}$Tb system. The solid and hollow triangles are the ICF cross 
sections corresponding to the $\alpha3n$ and $\alpha4n$ channels respectively. 
The dot-dot-dashed and the dashed lines are the uncoupled and 
coupled channels calculations performed with the code CCFULL. The solid line 
shows the coupled channels calculations scaled by the factor 0.86.\\
\\
Fig.3.  Complete and incomplete fusion cross sections (circle and
triangle) as a function of the centre-of mass energy for the 
$^{7}$Li+$^{159}$Tb system. The triangles show the ICF cross sections 
corresponding to the dominant $\alpha2n$ channel. The solid points are the 
present measurements and the hollow points are from Ref.[23]. 
The dot-dot-dashed and the dashed lines are the uncoupled and coupled channels 
calculations performed with the code CCFULL. The solid line shows the coupled 
channels calculations scaled by the factor 0.74.\\
\\
Fig.4.  Reduced complete fusion excitation functions for the 
$^{11,10}$B+$^{159}$Tb and $^{7}$Li+$^{159}$Tb systems.\\ 

\begin{thebibliography}{99}
\bibitem{Yo96} A. Yoshida {\em{et al.}}, Phys. Lett. B 389 (1996) 457.

\bibitem{Reh98} K.E. Rehm {\em{et al.}}, Phys. Rev. Lett. 81 (1998) 3341.

\bibitem{Ko98} J.J. Kolata {\em{et al.}}, Phys. Rev. Lett. 81 (1998) 4580.

\bibitem{Sig04} C. Signorini {\em{et al.}}, Nucl. Phys. A 735 (2004) 329.

\bibitem{Pie04} A. Di Pietro {\em{et al.}}, Phys. Rev. C 69 (2004) 044613

\bibitem{Dass93} C.H. Dasso, A. Vitturi, Phys. Rev. C 47 (1993) 2470(R).

\bibitem{Ha00} K. Hagino {\em{et al.}}, Phys. Rev. C 61 (2000) 037602.

\bibitem{Dia02} A. Diaz-Torres, I.J. Thompson, Phys. Rev. C 65 (2002) 024606.

\bibitem{Utso83} Utsonomiya {\em{et al.}} Phys. Rev. C 28 (1983) 1975

\bibitem{Zo78} D.R. Zolnowski {\em{et al.}}, Phys. Rev. Lett. 41 (1978) 92.

\bibitem{Bar80} J.H. Barker {\em{et al.}}, Phys. Rev. Lett. 43 (1980) 424.

\bibitem{Das99} M. Dasgupta {\em{et al.}}, Phys. Rev. Lett. 82 (1999) 1395.
                                                                                
\bibitem{Sig99} C. Signorini {\em{et al.}}, Eur. Phys. J A5 (1999) 7.

\bibitem{Das02} M. Dasgupta {\em{et al.}}, Phys. Rev. C 66 (2002) 041602(R).

\bibitem{Das04} M. Dasgupta {\em{et al.}}, Phys. Rev. C 70 (2004) 024606 and
and references therein.
                                                                                
\bibitem{Tri02} V. Tripathi {\em{et al.}}, Phys. Rev. Lett. 88 (2002) 172701;
Phys. Rev. C 72 (2005) 017601.

\bibitem{Mor00} S.B. Moraes {\em{et al.}}, Phys. Rev. C 61 (2000) 064608.

\bibitem{Dia03} C. Beck {\em{et al.}}, Phys. Rev. C 67 (2003) 054602.

\bibitem{Mu1} A. Mukherjee {\em{et al.}}, Phys. Lett. B 526 (2002) 295.

\bibitem{Mu2} A. Mukherjee {\em{et al.}}, Nucl. Phys. A 596 (1996) 299.\\
              A. Mukherjee {\em{et al.}}, Nucl. Phys. A 635 (1998) 205.
                                                                                
\bibitem{Mu3} A. Mukherjee {\em{et al.}}, Nucl. Phys. A 645 (1999) 13.

\bibitem{Mu4} A. Mukherjee and B. Dasmahapatra, Phys. Rev. C 63 (2000) 017604.

\bibitem{Bro75} R. Broda {\em{et al.}}, Nucl. Phys. A 248 (1975) 356.

\bibitem{Gav80} A. Gavron, Phys. Rev. C 21 (1980) 230.

\bibitem{Hag99} K. Hagino {\em{et al.}}, Comput. Phys. Commun. 123 (1999) 143.

\bibitem{Brog81} R.A. Broglia and A. Winther, {\em{Heavy Ion Reactions}},
Vol. 1 (Benjamin/Cummings, Reading, MA, 1981).

\bibitem{New1} J.O. Newton {\em{et al.}}, Phys. Lett. B 586 (2004) 219.

\bibitem{Mu5} A. Mukherjee {\em{et al.}}, to be communicated.

\bibitem{Mu6} A. Mukherjee {\em{et al.}}, Phys. Rev. C 66 (2002) 034607.

\bibitem{Ram87} S. Raman {\em{et al.}}, At. Data Nucl. Data Tables 36 (1987) 1.

\bibitem{Mol95} P. M$\ddot{\mbox{o}}$ller {\em{et al.}}, At. Data Nucl. Data 
Tables 59 (1995) 185.
\end{thebibliography}
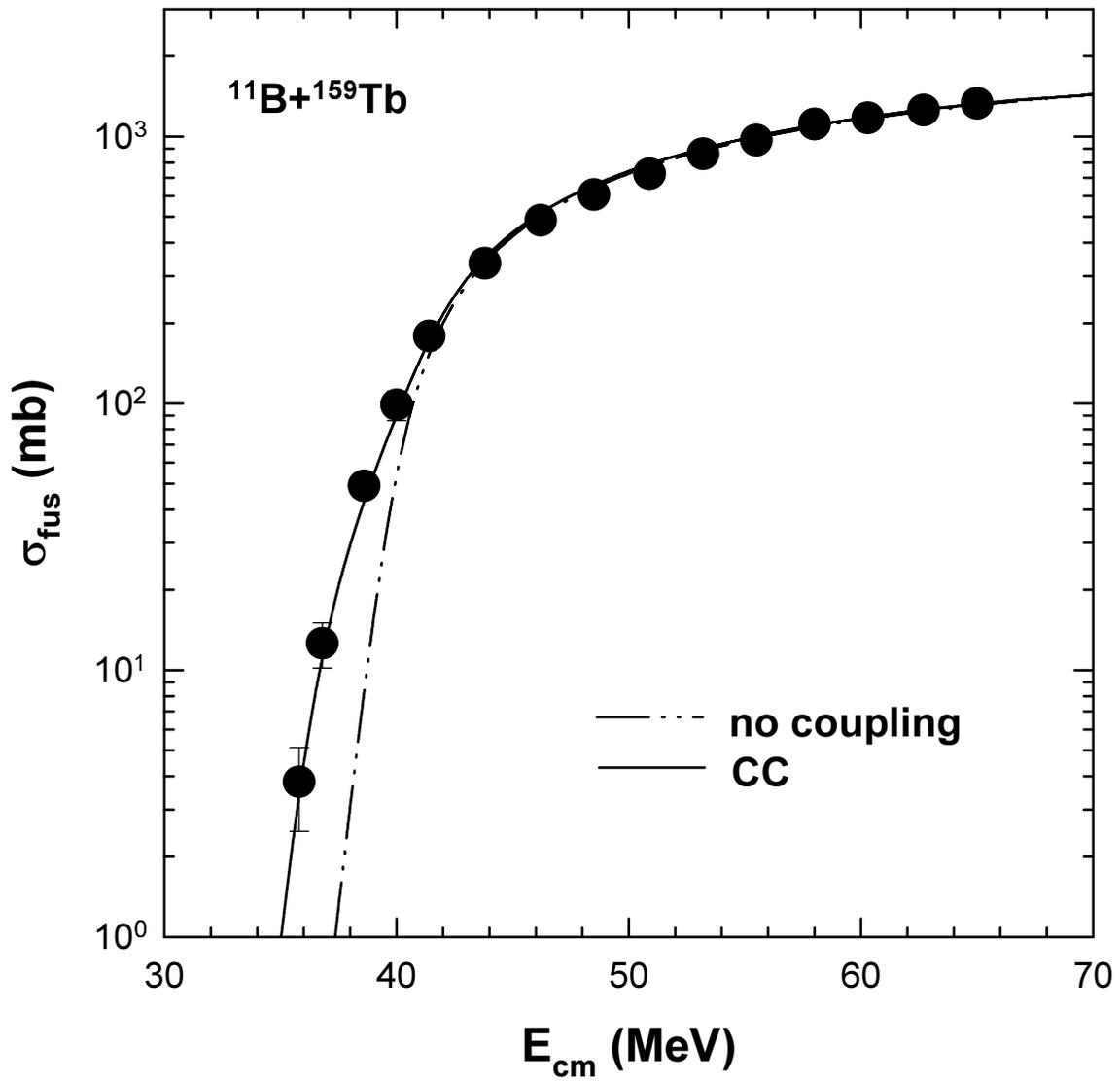
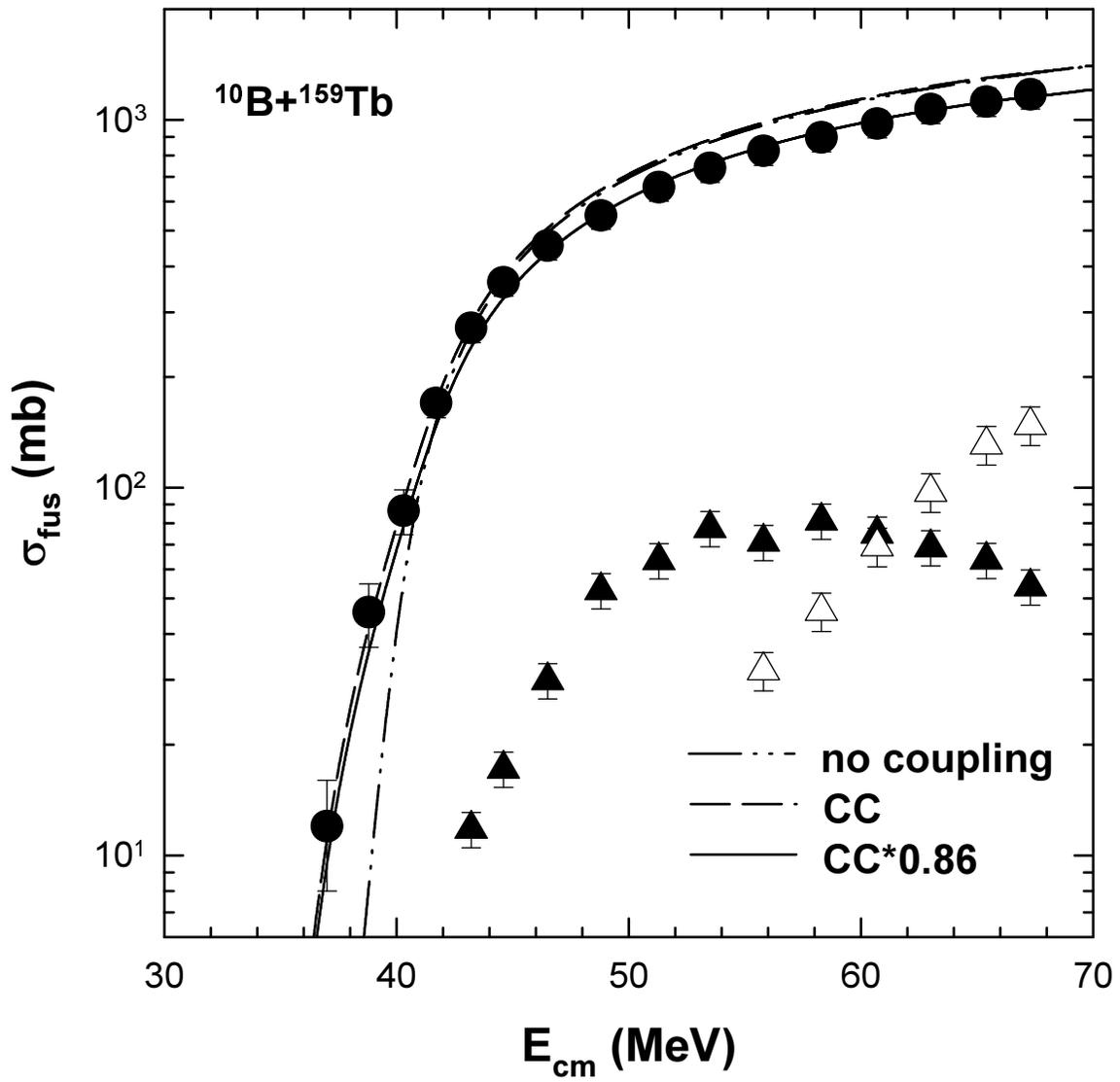
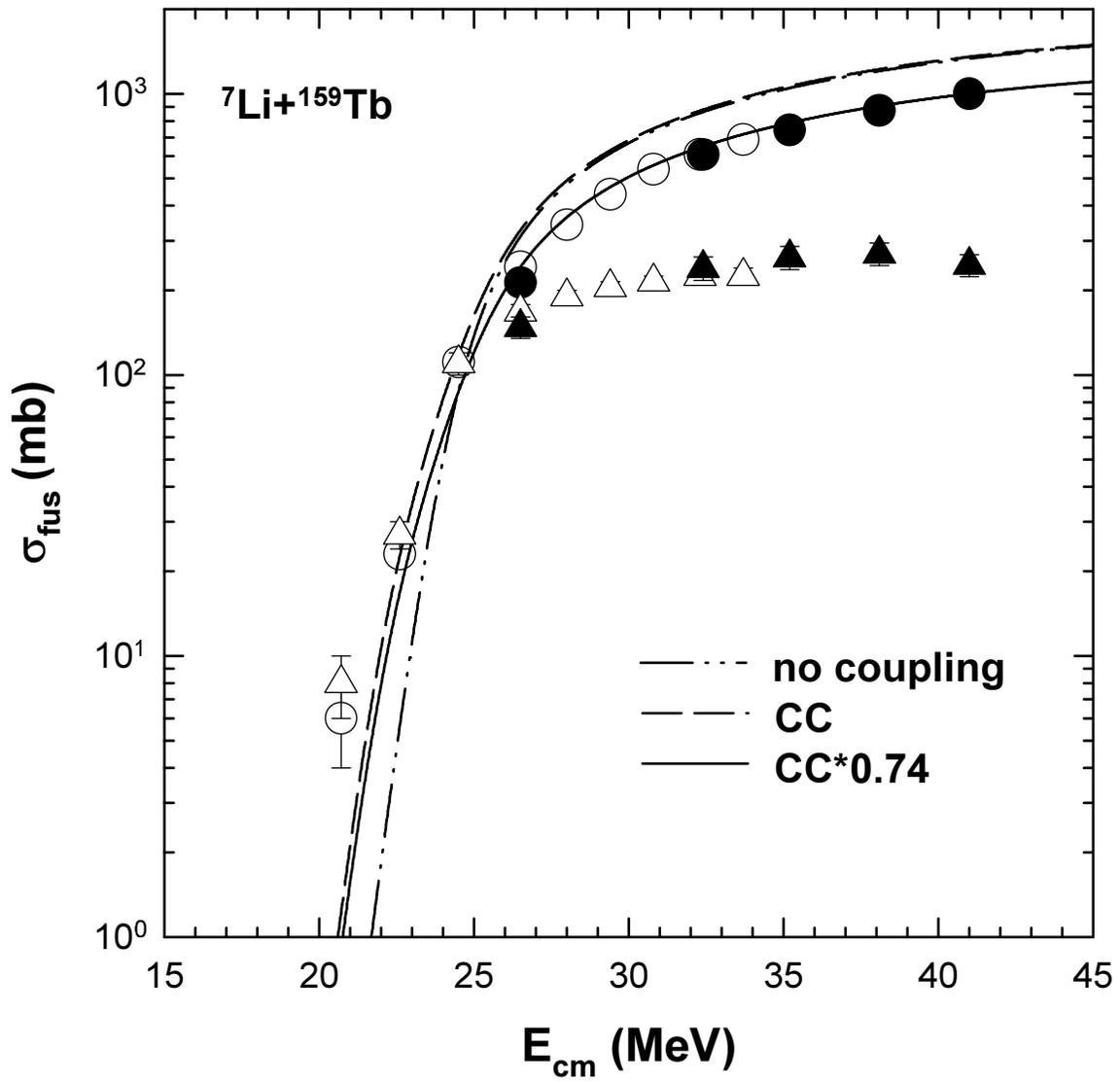
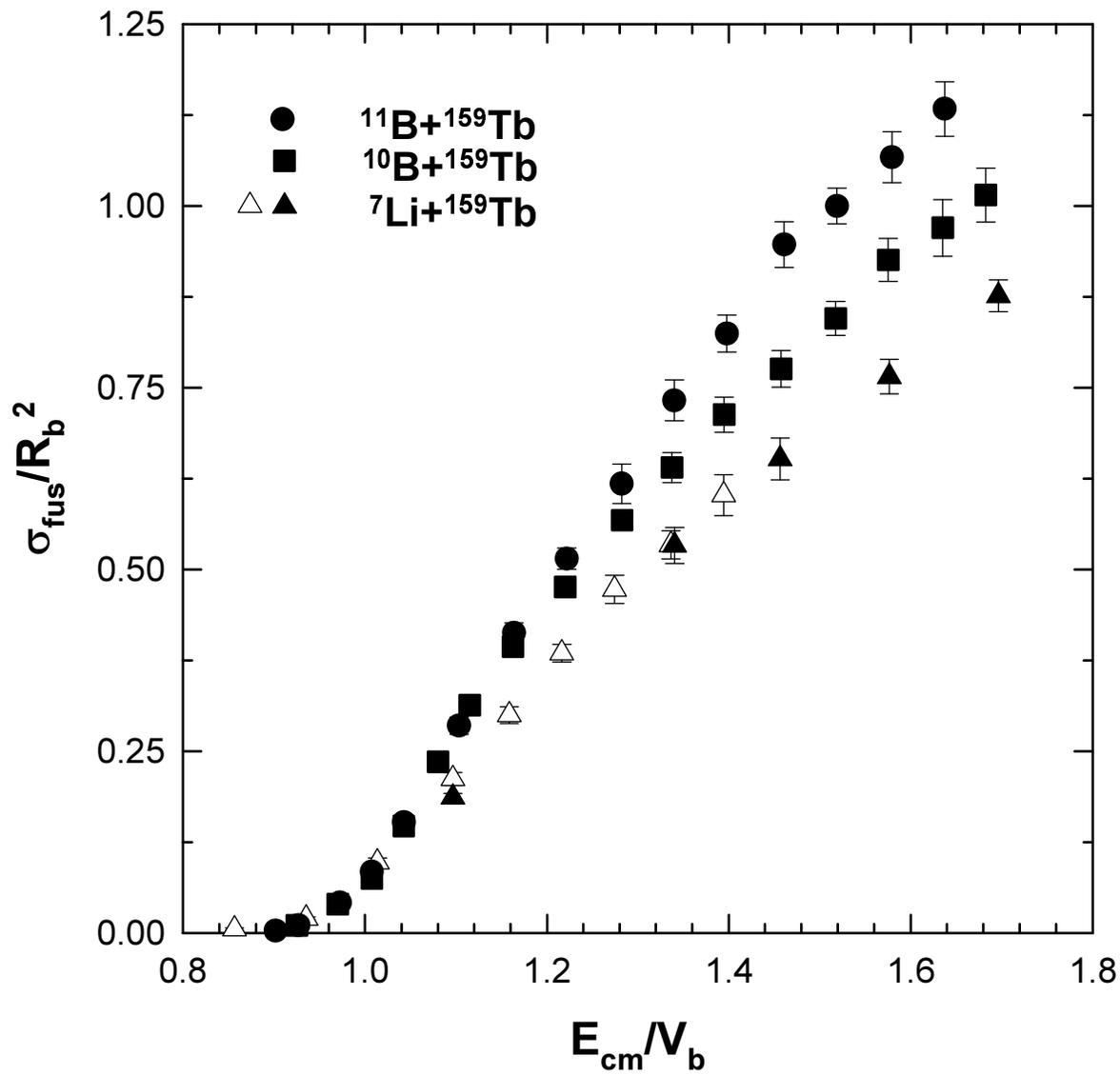
\end{document}